\documentclass[a4paper,fleqn,usenatbib]{mnras}

\usepackage[T1]{fontenc}
\usepackage{ae,aecompl}

\usepackage{graphicx}	
\usepackage{amsmath}	
\usepackage{amssymb}	

\usepackage{natbib}

\usepackage{latexsym,amssymb}
\usepackage{ucs}
\usepackage[utf8x]{inputenc}
\usepackage{color}

\newcommand{\Msun}{M$_\odot$}

\title[Rotation of NSCs formed by cluster-inspirals]{On the rotation of nuclear star clusters formed by cluster-inspirals}

\author[A. Tsatsi et al.]{Athanasia Tsatsi,$^{1}$\thanks{E-mail: tsatsi@mpia.de}
Alessandra Mastrobuono-Battisti,$^{1}$
Glenn van de Ven,$^{1}$
\newauthor
Hagai B. Perets,$^{2}$
Paolo Bianchini$^{1}$
and Nadine Neumayer$^{1}$
\\
$^{1}$Max-Planck-Institut f{\"u}r Astronomie, K\"onigstuhl 17, 69117 Heidelberg, Germany\\
$^{2}$Physics Department, Technion - Israel Institute of Technology, Haifa, Israel 32000\\
}

\date{Accepted XXX. Received YYY; in original form ZZZ}

\pubyear{2016}

\begin{document}
\label{firstpage}
\pagerange{\pageref{firstpage}--\pageref{lastpage}}
\maketitle

\begin{abstract}
Nuclear Star Clusters (NSCs) are commonly observed in the centres of most galactic nuclei, including our own Milky Way. While their study can reveal important information about the build-up of the innermost regions of galaxies, the physical processes that regulate their formation are still poorly understood. NSCs might have been formed through gas infall and subsequent in situ star formation, and/or through the infall and merging of multiple star clusters into the centre of the galaxy. Here, we investigate the viability of the latter, by studying direct N-body simulations of inspiralling clusters to the centre of a Milky-Way-like nuclear bulge that hosts a massive black hole. We find that the NSC that forms through this process can show both morphological and kinematical properties that make it comparable with observations of the Milky Way NSC, including significant rotation-- a fact that has been so far attributed mainly to gas infall. We explore its kinematic evolution, to see if and how the merger history can imprint fossil records on its dynamical structure. Moreover, we study the effect of stellar foreground contamination in the line-of-sight kinematics of the NSC. Our study shows that no fine tuning of the orientation of the infalling globular clusters is necessary to result in a rotating NSC. We suggest that cluster-inspiral is a viable mechanism for the formation of rotating NSCs.
\end{abstract}




\section{Introduction}
\label{sec:intro}

Nuclear Star Clusters (NSCs) are massive and compact stellar clusters found in the central regions of most galaxies. With half-light radii of a few parsecs \citep[e.g.][]{Georgiev_2014} and typical dynamical masses of \ensuremath{10^{ 6 }-10^{ 7 } $\Msun$ }, they are thought to be the densest stellar systems in the Universe \citep{Walcher_2005}. The nearest NSC that can be observed lies within the central 10 pc of our own Galaxy, with a half-light radius of 4.2 $\pm$ 0.4 pc \citep{Schoedel_2014} and a mass of \ensuremath{2-3 \times10^{ 7 } $\Msun$ } \citep{Schoedel_2014, Feldmeier_2014}.

Over the last decade, a series of studies have shown that NSCs are extremely common: more than 77\% of late type galaxies host a NSC at their centre \citep{Boeker_2002, Georgiev_2014}, as well as at least 66\% of early-type galaxies, mainly dwarf ellipticals and lenticulars \citep{Cote_2006, Turner_2012, denBrok_2014}. Those fractions are only a lower limit to the true fraction of galaxies hosting NSCs, mainly due to several observational biases that limit their detection across the Hubble sequence.

NSCs seem to correlate with global properties of their host galaxies.  It has been suggested that the masses of NSCs and the masses of supermassive black holes (SMBHs) share similar-slope correlations with the host stellar velocity dispersion and bulge luminosity \citep{Wehner_2006, Rossa_2006, Ferrarese_2006}. However, a number of  studies have questioned this similarity \citep[e.g.][]{Graham_2012, Erwin_2012, Scott_2013, Kormendy_Ho_2013}. Although it is not clear if the formation and growth mechanisms of NSCs are coupled to that of SMBHs, they both seem to be connected with the evolution of their host galaxies.

Two main scenarios have been proposed to explain the formation of NSCs; 1) the in situ formation model \citep{Loose_1982, Milosavljevic_2004}, according to which the NSC forms as gas infalls to the centre of the galaxy, where subsequently star formation takes place locally and most likely in an episodic manner \citep[e.g.][]{Schinnerer_2008}, and 2) the cluster-inspiral scenario \citep{Tremaine_1975, Capuzzo_Dolcetta_1993, Antonini_2012, Gnedin_2014}, where the NSC is formed by the accretion of globular clusters, that infall to the centre due to dynamical friction. Both of these models can explain the mixture of stellar populations of different ages in NSCs \citep[e.g.][]{Walcher_2006, Rossa_2006, Seth_2010}. Until now, it is not clear, which model works best to explain the observations, or whether both of these processes are working in parallel to form and grow NSCs.

The detailed study of the dynamical properties of NSCs can, however, provide an important tool to disentangle the possible formation mechanisms of NSCs and understand the relative importance of each mechanism, in case they work in parallel. 

The kinematics of NSCs formed via the cluster-inspiral scenario has so far been focused on simulations tuned for extragalactic NSCs \citep{Hartmann_2011}. They found that NSCs formed solely through this mechanism can not exhibit the high amount of rotation that is actually observed. Additionally they show a central peak in their second order kinematic moment $\ensuremath{V_{RMS}=\sqrt{V^2+\sigma^2}}$ that is too high to agree with observations.

Although this might be the case for some extragalactic NSCs, in the case of the Milky Way (MW) NSC, recent findings by \cite{Feldmeier_2014} show a central peak in $V_{RMS}$, as well as strong evidence for a polar kinematic substructure in its central region and a kinematic misalignment between the main body of the NSC and the Galactic plane. These observations give evidence that globular cluster inspirals may indeed play an important role in the main build-up process of the MW NSC. However it is not yet clear if and how this mechanism can account for the rotation observed in NSCs, which is often attributed to gas infall, or star cluster infall from the galactic disc \citep[e.g.][]{Seth_2008}.

Here, we investigate further if and how the cluster-inspiral scenario can reproduce the observed properties of NSCs, focusing in particular on the kinematic signatures of the MW NSC. We use a set of N-body simulations of the formation of a NSC through the consecutive infall of globular clusters (GCs) in a Milky Way-like nuclear bulge with a central massive black hole (MBH) \citep{Antonini_2012, Perets_2014}. The NSC is analyzed in an observational-like manner, constructing mock photometric and line-of-sight stellar kinematic maps that we then use to assess the dynamical properties of the simulated NSC. 

This paper is organised as follows: in Section~\ref{sec:Simulations} we describe the N-body simulations, in Section~\ref{sec:Evolution} we show the kinematic evolution of the NSC and in Section~\ref{sec:Kinematics} we compare the kinematic and morphological properties of the simulated NSCs to those of the MW NSC. In Section~\ref{sec:Contamination} we discuss the effect of the contamination due to the nuclear bulge and, finally, we conclude in Section~\ref{sec:Summary}.

\section{Simulations}
\label{sec:Simulations}

The N-body simulations used in this work are described in detail in \cite{Antonini_2012} and \cite{Perets_2014}. They simulate the formation of a Milky Way-like NSC through the consecutive infall of 12 identical globular clusters (GCs) with a mass of \ensuremath{1.1 \times10^{ 6 } $\Msun$ } each, in the inner region of a nuclear bulge ($M_{nb}=$ \ensuremath{10^{ 8 } $\Msun$ }), hosting a central MBH ($M_{\bullet}=$ \ensuremath{4 \times10^{ 6 } $\Msun$ }), similar to the MW MBH \citep{Genzel_2010}. Each GC is represented by a tidally truncated \cite{King_1966} model and is initially moving on a circular orbit with randomly chosen parameters, at a galactocentric distance of 20 pc. The time interval is kept constant between infalls and is $\sim$0.85 Gyr, rescaled to the real mass of the particles, as described by \cite{Mastrobuono_2013}. After the last infall, the NSC and the surrounding bulge are let to evolve in isolation for $\sim$2.2 Gyr, adding up to a total simulation time of $\sim$12.4 Gyr. The total mass of the resulting NSC is approximately \ensuremath{1.4 \times10^{ 7 } $\Msun$ }. This value is in agreement on the 2-sigma level with the mass of the MW NSC (\ensuremath{2.5 \pm 0.6 \times10^{ 7 } $\Msun$ }), as estimated by \cite{Schoedel_2014}.

Here we analyse three realisations of the initial conditions described above, with different randomisations of the initial orbital parameters of the infalling GCs (see Table~\ref{tab:a} for the orbital parameters used in each simulation). In all simulations the longitude of ascending node $\Omega$ and inclination i of the GC orbit are randomly chosen. Simulation III differs from Simulations I and II because i is chosen with the constraint that i $\textless 90^{\circ}$, so that the GCs infall with a similar orbital direction to the centre of the Galaxy (only prograde orbits). This choice of initial parameters has been made to represent clusters that might have initially formed in the central molecular zone of the MW, that at 20 pc distance from the centre will have random offsets with respect to the Galactic plane, but all share a similar orbital spin.

\begin{table}

\begin{center}
\caption{Initial orbital parameters of the 12 infalling clusters: longitude of ascending node $\Omega$ and inclination i are given for the three simulations, calculated with respect to the same simulation reference frame.}
\begin{tabular}{rcccccc}
\hline\hline
n & \multicolumn{2}{c}{Simulation I} & \multicolumn{2}{c}{Simulation II} & \multicolumn{2}{c}{Simulation III}  \\
&$\Omega$ &$i$& $\Omega$ & $i$ & $\Omega$ & $i$\\
&(deg)&(deg)&(deg)&(deg)&(deg)&(deg)\\
\hline
1&82.4&60.7&171.3&118.6	& 175.9	&10.0	\\
2&327.7 &	178.7&237.8	 &173.0	&7.8	&35.0	\\
3&76.2 & 139.5	& 325.9	&143.0	&284.7&6.7	\\
4&290.6 & 171.3&39.6 	&26.9	&314.1&20.3	\\
5&335.4 & 24.6	&89.3 	&117.7	&224.9&23.0	\\
6&300.6 & 18.2	& 27.9	&9.5		&254.8	&10.7	\\
7&343.9 & 173.9& 232.9	&6.3		&246.7	&39.9	\\
8&47.9 & 	128.9& 262.5	&22.2	&126.2	&87.8	\\
9&272.0 & 2.3	& 51.1	&174.1	&326.8&7.1	\\
10&41.3 & 139.0& 316.6	&94.8	&52.2&79.8	\\
11&300.9 & 153.5& 165.9	&4.2		&9.1	&29.4	\\
12&318.2 & 120.2& 61.8	&79.1	&136.4	&35.7	\\

\hline
\end{tabular}
\label{tab:a}
\end{center}

\end{table}

\section{Kinematic evolution of the NSC}
\label{sec:Evolution}

\begin{figure}
\center
\includegraphics[trim = 3.1cm 13.8cm 0cm 3.8cm,clip, width=9.5cm, ]{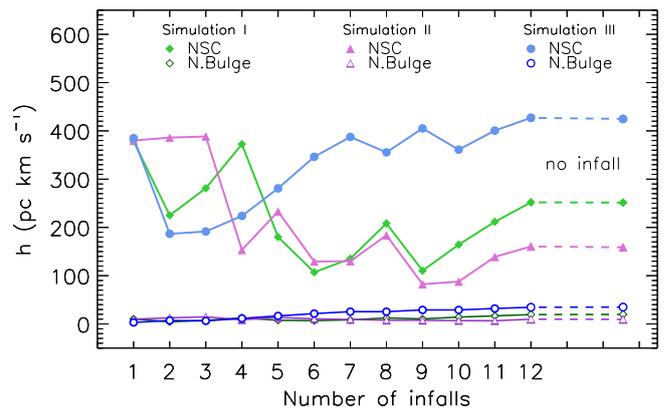}
\center
\caption{Evolution of the specific angular momentum h of the simulated NSCs after each infall. The green (diamonds), magenta (triangles) and blue (circles) points correspond to Simulation I, II, and III, respectively. Filled symbols correspond to the NSC, while open symbols correspond to the surrounding nuclear bulge particles. Dashed lines correspond to the last timestep of the simulation where the system evolves in isolation (no infall).}
\label{fig:figure1}
\end{figure}

Figure~\ref{fig:figure1} shows the evolution of the specific angular momentum (i.e. the total angular momentum divided by mass) of both the NSC remnant and the surrounding nuclear bulge after each infall for the three simulation set-ups. The NSC remnants show strong angular momentum variance after each infall, which depends on the orientation of the infalling GC. Finally after the 12th infall, the system is let to evolve in isolation, resulting in no change in angular momentum. All three NSC remnants show strong rotation in the final snapshots as can be seen from their high specific angular momentum. Simulation III results in a NSC with significantly higher angular momentum, as a consequence of the missing GCs in retrograde orbits. 
Figure~\ref{fig:figure1}  also shows the evolution of the specific angular momentum of the surrounding bulge for all simulations. The nuclear bulge rotation shows a small increase with time. The amount of nuclear bulge rotation is larger for more strongly rotating NSCs, meaning that the kinematics of the nuclear bulge can be affected by the NSC formation mechanism. The amount of nuclear bulge rotation in the final snapshot is, however, very low for all the three models.

\begin{figure}
\center
\includegraphics[trim = 0.8cm 11cm 0.2cm 3.2cm,clip, width=9.cm, ]{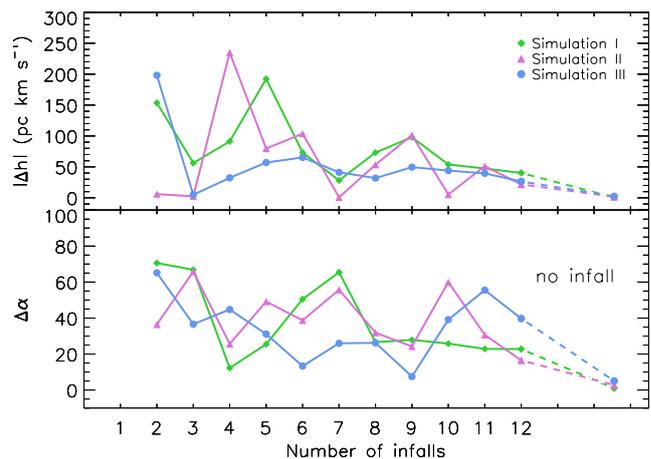}
\center
\caption{\textit{Top:} Change of the specific angular momentum $|\Delta h|$ of the NSC after each infall. The green (diamonds), magenta (triangles) and blue (circles) points correspond to Simulation I, II, and III, respectively. \textit{Bottom:} Evolution of the precession angle $\Delta \alpha$ of the angular momentum vector of the NSC after each infall. Note the anti-correlation of  $|\Delta h|$ and $\Delta \alpha$ for each simulation.}
\label{fig:figure2}
\end{figure}

Figure~\ref{fig:figure2}  shows the evolution of $\lvert\Delta h\rvert$, the specific angular momentum variation of the NSC after each infall, as well as the evolution of the precession angle $\Delta \alpha$ of the NSC. The latter shows the change in the orientation of the angular momentum vector (rotation axis) of the NSC after each infall.

 The variation of $\lvert\Delta h\rvert$ is higher during the first infalls and becomes gradually smaller during the last infalls, as the NSC grows in mass and consequently the mass ratio between every infalling GC and the growing NSC becomes smaller. The specific angular momentum of the NSC shows no variation during the last $\sim$2 Gyr of evolution where there is no infall. 

The precession angle $\Delta \alpha$ of the NSC on the other hand, seems to show a more stochastic evolution. Even after the last infall, where the mass ratio of the inspiralling GC and the growing NSC is as small as 1:11, the precession angle ranges from $15^{\circ}$ (Simulation II) to as much as $40^{\circ}$ (Simulation III), depending on the inclination of the infalling GC and the dynamical structure of the NSC. When the system is let to evolve in isolation, the angular momentum vector shows almost no precession for Simulations I, II, while  $\Delta \alpha \sim 5^{\circ}$ for Simulation III.

Note the anti-correlation between $\lvert\Delta h\rvert$ and $\Delta \alpha$, at least for the first $\sim 9$ infalls in all simulations. This anti-correlation reflects the way the orbit inclination of an infalling cluster impacts the resulting angular momentum of the growing NSC. A prograde or retrograde infall will cause a high $\lvert\Delta h\rvert$ and low $\Delta \alpha$, while infalls of intermediate inclinations will have the opposite effect. The anti-correlation becomes weaker as the growing NSC becomes more massive (after the 9th infall).

We note that even if the globular clusters are initially randomly distributed around the center, their net angular momentum is not negligible. Most of their net angular momentum is transferred to the NSC. For example, in Simulation I, $\sim65\%$ of the total input angular momentum of the system is transferred to the NSC in the final timestep. The rest $\sim35\%$ is transferred to the surrounding nuclear bulge, which, however, does not correspond to a high rotation, as shown in Figure~\ref{fig:figure1}.

\section{Kinematics and morphology of the NSC}
\label{sec:Kinematics}

\begin{figure*}
\center
\includegraphics[trim = 2.cm 12.8cm 2cm 6cm,clip, width=16cm, ]{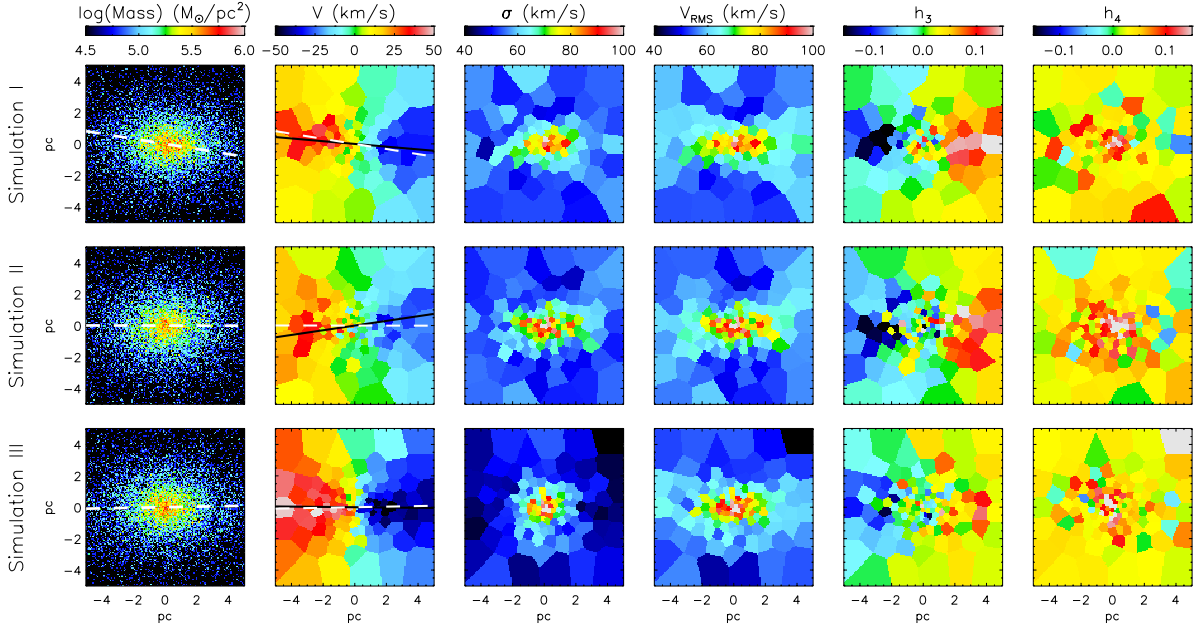}
\center
\caption{LOSVD of the simulated NSC. From left to right: Projected stellar mass surface density, line-of-sight velocity v, velocity dispersion $\sigma$ in km$\cdot$s$^{-1}$, and higher-order moments $h_3$ and $h_4$, comparable to the skewness and the kurtosis, respectively. The white dashed line shows the major photometric axis, while the solid black line shows the kinematic major axis of each cluster.}
\label{fig:figure3}
\end{figure*}

\begin{figure*}
\center
\includegraphics[trim = 2.65cm 13.3cm 2.5cm 0.7cm,clip, width=11cm, ]{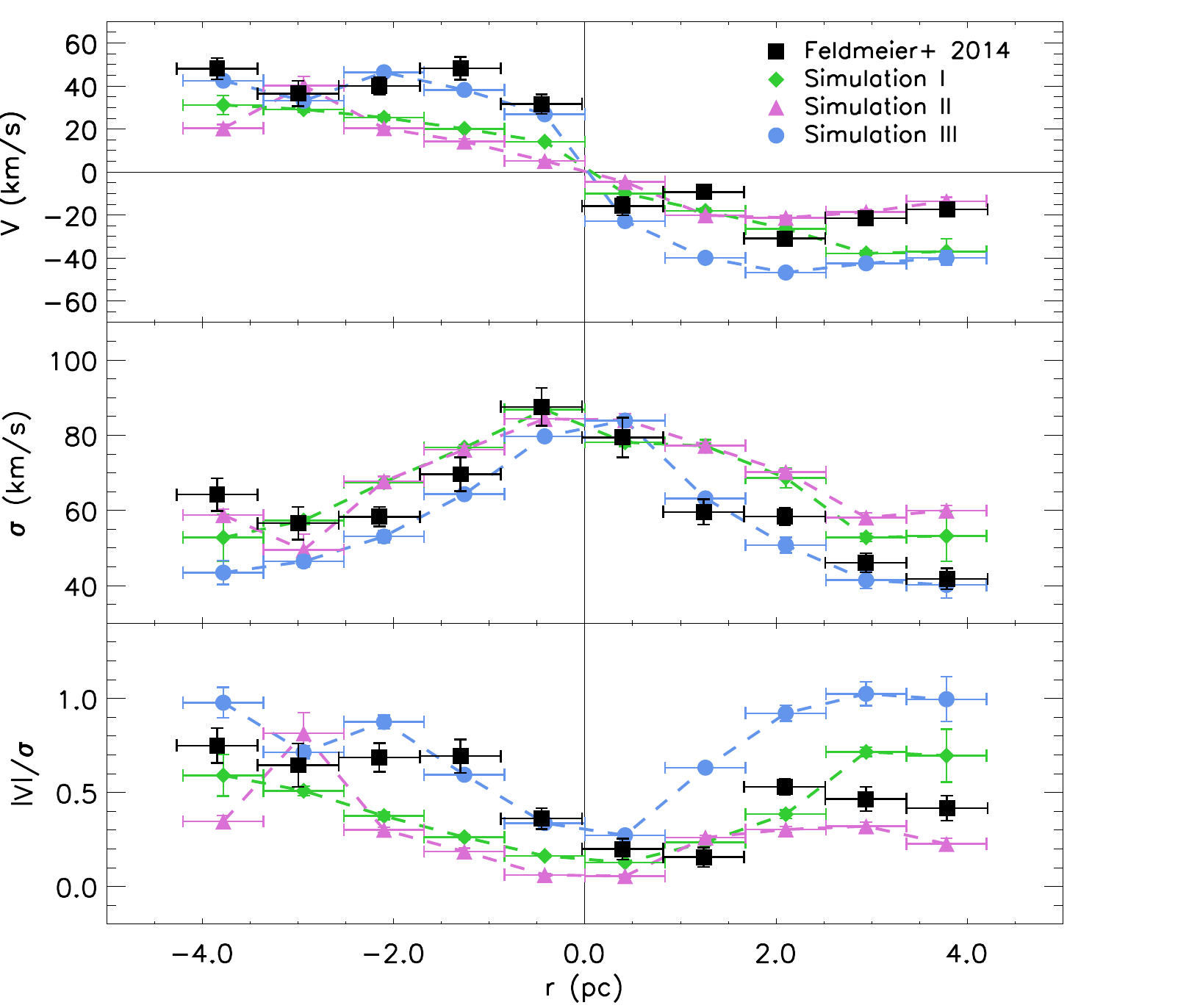}
\center
\caption{Kinematic profiles ($V$, $\sigma$ and $V/\sigma$) for the three simulated clusters (dashed lines) compared to the corresponding profiles of the Milky Way NSC (black squares) by \protect\cite{Feldmeier_2014}. All profiles are extracted from a slit along the kinematic axis. The asymmetry between left and right side of the MW NSC is caused by dust extinction \protect\citep{Chatzopoulos_2015b}.}
\label{fig:figure4}
\end{figure*}

\subsection{Mock kinematics}
\label{sec:mock_kin}

In order to compare the orbital and mass distribution of the final NSCs with observable properties, we create two-dimensional mock stellar mass and kinematic maps as follows. Particles are projected along a line-of-sight which is perpendicular to the total angular momentum vector of the NSC, meaning that the line-of-sight rotation observed should be maximum. Particles are then binned on a regular grid centred on the centre of mass of the cluster, with a field-of-view (FoV) of 10 pc$\times$10 pc and a pixel size of 0.08 pc.

 The bulk velocity of the NSC is estimated within a sphere of 50 pc around the centre and subtracted from all particle velocities. The extracted kinematic maps are spatially binned using the 2D Voronoi binning method \citep{Cappellari_2003}, based on a minimum number of particles per pixel in the map. Signal corresponds to the number of particles per pixel and we adopt Poisson noise, such that our signal-to-noise ratio per bin (\ensuremath{SN_{bin}}) corresponds approximately to a target value $\ensuremath{SN_{T}\sim15}$.

  The stellar line-of-sight velocity distribution (LOSVD) is then extracted and fitted with the Gauss--Hermite series \citep{vdMarel_1993}, as implemented by \citet[Appendix]{vdVen_2006}, which allows us to extract the Gauss--Hermite parameters for every bin (V, $\sigma$, $h_3$ and $h_4$). The mass and stellar LOSVD of the three simulated NSCs are shown in Figure~\ref{fig:figure3}.

\subsection{Comparison with observations of the MW NSC}

In order to compare with observations, we choose a FoV of 5 pc radius, which is approximately the half-light radius of the MW NSC \citep{Schoedel_2014}\footnote{We note however, that \cite{Fritz_2016} report values for the half-light radius of the MW NSC that range from 5 to 9 pc.}. The half-mass radius of our simulated NSC is approximately 10 pc for all simulation set-ups. We would expect differences between observed half-light and half-mass radius of the MW NSC if the mass-to-light ratio is not constant, as a result of the non-trivial interplay between mass segregation and the presence of young bright stars in the central region \citep[e.g.][]{Paumard_2006}. Within 5pc, the simulated NSC matches the observed shape of the surface density distribution of the MW NSC \citep{Antonini_2012}. Therefore, we limit our kinematic analysis and comparison to this radial extent.
Using the first and second moments of the intensity distribution of our mock images, we find the position of the projected major axis and the flattening $q=b/a$ of our simulated NSCs within the adopted FoV of 5 pc radius. The average flattening of the NSC is $q=0.64$ for Simulation I, and $q=0.69$ for Simulations II and III. These values are remarkably similar to the observed flattening of the MW NSC, $q_{obs}=0.71\pm0.02$ \citep{Schoedel_2014}.

The NSC shows a significant amount of rotation, of an amplitude of $\sim$40 km$\cdot$s$^{-1}$ within 5 pc for Simulation I and II. The velocity is higher ($\sim$50 km$\cdot$s$^{-1}$) for Simulation III, where the infalling GCs have a similar initial orbital direction. 

In order to compare our results with the observed kinematic profiles of the MW NSC, we estimate the kinematic major axis of the NSC within the adopted FoV using the kinemetry method, as developed by \cite{Krajnovic_2006}. The kinematic axis for each simulated NSC is shown in Figure~\ref{fig:figure3}  (solid black lines). We then place a mock slit along the kinematic axis, of width of 0.84 pc and extract the LOSVD of the simulated clusters in equal-size bins of 0.84 pc size, which corresponds to a binning similar to the one used by \cite{Feldmeier_2014} to the MW NSC. The corresponding errors are calculated by Monte Carlo simulations of the extracted LOSVD \citep[see][]{vdVen_2006}. The profiles of $V$, $\sigma$ and $V/\sigma$ for the three simulations are shown in Figure~\ref{fig:figure3}. The kinematic profiles show a very good agreement with the kinematic profiles observed in the MW NSC \citep{Feldmeier_2014}.

\subsection{Kinematic misalignments}

Figure~\ref{fig:figure3} shows the measured kinematic and the photometric major axes of all simulated NSCs within the adopted FoV. We find that the offset between these two axes within 5pc is $\Delta\theta\sim4.2^{\circ}$, $8.6^{\circ}$ and $0.5^{\circ}$ for Simulations I, II, and III, respectively. Simulation I also shows a misalignment of about $9.2^{\circ}$ between the photometric major axis within 5pc and the projected plane, which is perpendicular to the total angular momentum vector of the NSC (the x axis of Figure~\ref{fig:figure3}). Simulation III, however, characterised by inspiralling GCs with similar orbital directions, shows no significant offset between the kinematic and the photometric axis of the resulting NSC.

Such a misalignment between kinematics and morphology has also been recently observed in the MW NSC, with a median value of $\Delta\theta\sim9^{\circ}\pm3^{\circ}$, suggesting this as an evidence that cluster-inspirals may have played an important role in the formation of the MW NSC \citep{Feldmeier_2014}. Here we confirm that this scenario is able to produce observable misalignments between the photometry and kinematics of the resulting NSCs, which are stronger in the case where the infalling GCs are employed in random orbital directions (Simulations I and II), however not in the case where the GCs infall with a similar orbital direction (Simulation III).
\begin{figure*}
\center
\includegraphics[trim = 3.8cm 7.5cm 0cm 5.cm,clip, width=19.7cm, ]{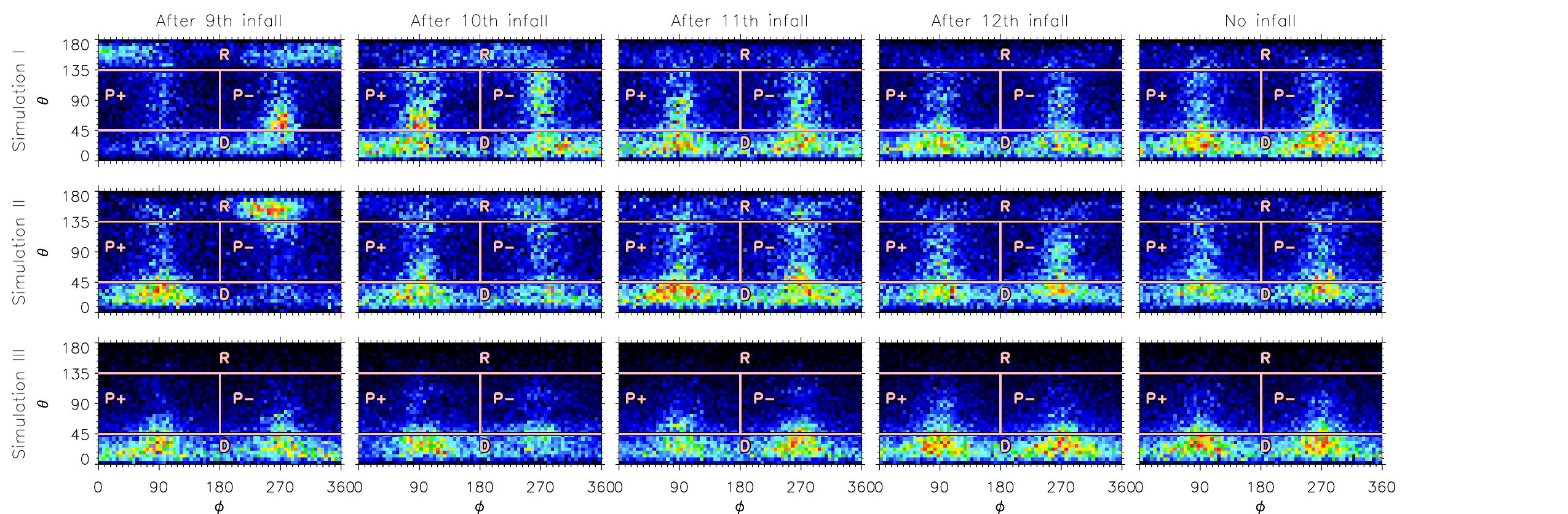}
\center
\caption{Evolution of the angular momentum vector distribution within a field-of-view of 10$\times$10 pc$^2$ for Simulations I, II, III (top to bottom) for the last 5 timesteps of each simulation (left to right). The angles $\theta$ and $\phi$ are the polar and azimuthal angles, respectively, of the angular momentum vector in a spherical coordinate system where the z-axis is the rotation axis of the NSC. The distribution is colour-coded according to the magnitude of the total angular momentum vector per pixel, normalised to the maximum value for every timestep. The distribution is divided in 4 regions that correspond to particles with prograde (D), retrograde (R) and polar spins (P+, P-), respectively.}
\label{fig:figure5}
\end{figure*}

\begin{figure}
\center
\includegraphics[trim = 0cm 0.cm 6cm 20.7cm,clip, width=8.5cm, ]{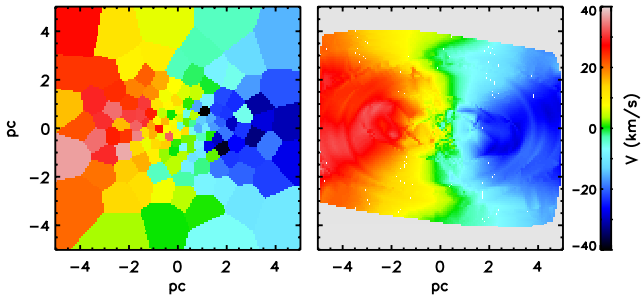}
\center
\caption{Left: Mock line-of-sight velocity map of the NSC of Simulation II, after the infall of the 12th GC. Right: the corresponding kinemetric model, showing a weak polar twist at $\sim$3 pc.}
\label{fig:figure6}
\end{figure}

\begin{figure*}
\center
\includegraphics[trim = 1.8cm 13.8cm 2cm 7cm,clip, width=16cm,]{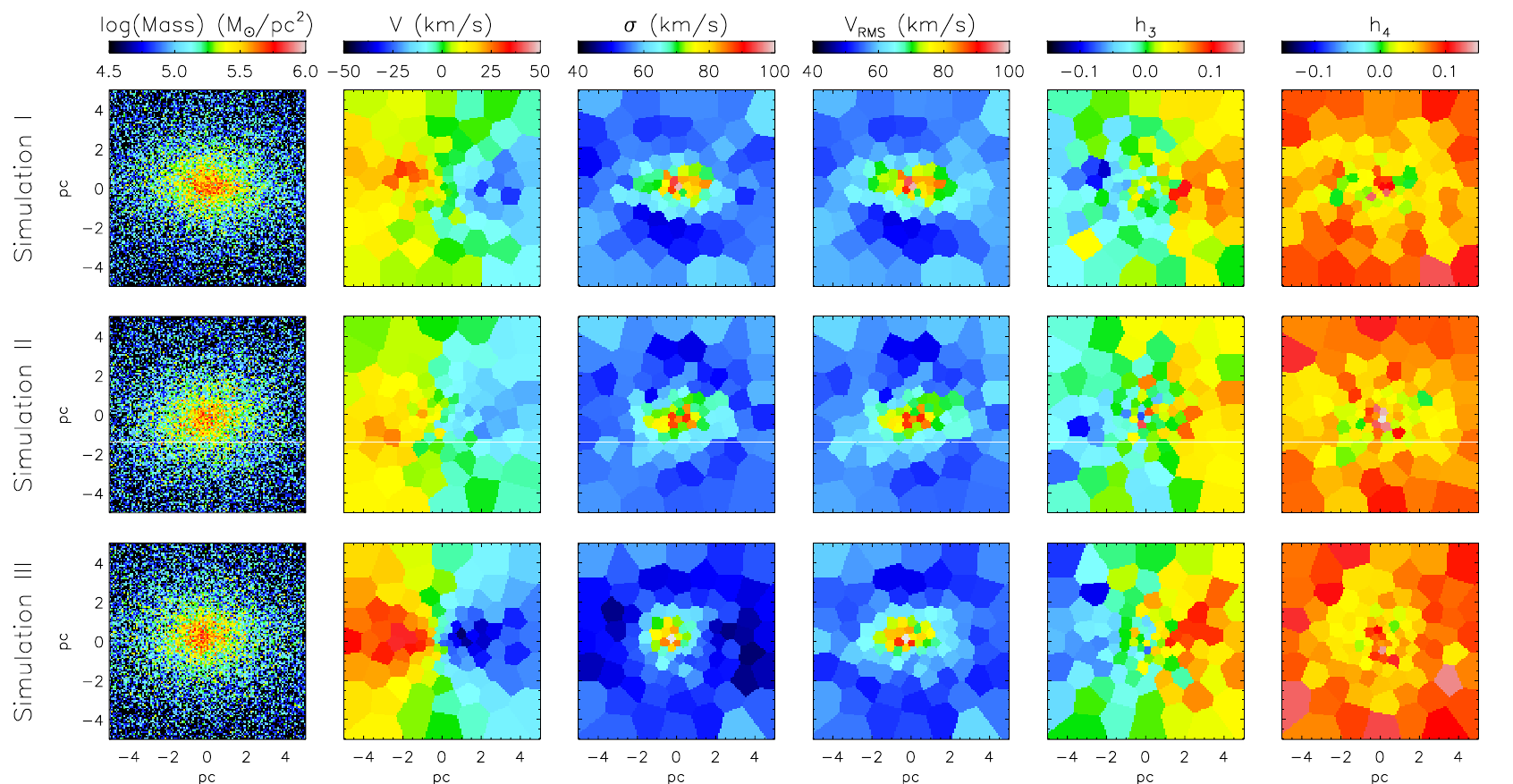}
\center
\caption{Same as Figure~\ref{fig:figure3}, where the projected NSC is contaminated with stars from the surrounding nuclear bulge.}
\label{fig:figure7}
\end{figure*}

\begin{figure}
\center
\includegraphics[trim = 2.65cm 13.4cm 2.5cm 0.7cm,clip, width=9.4cm, ]{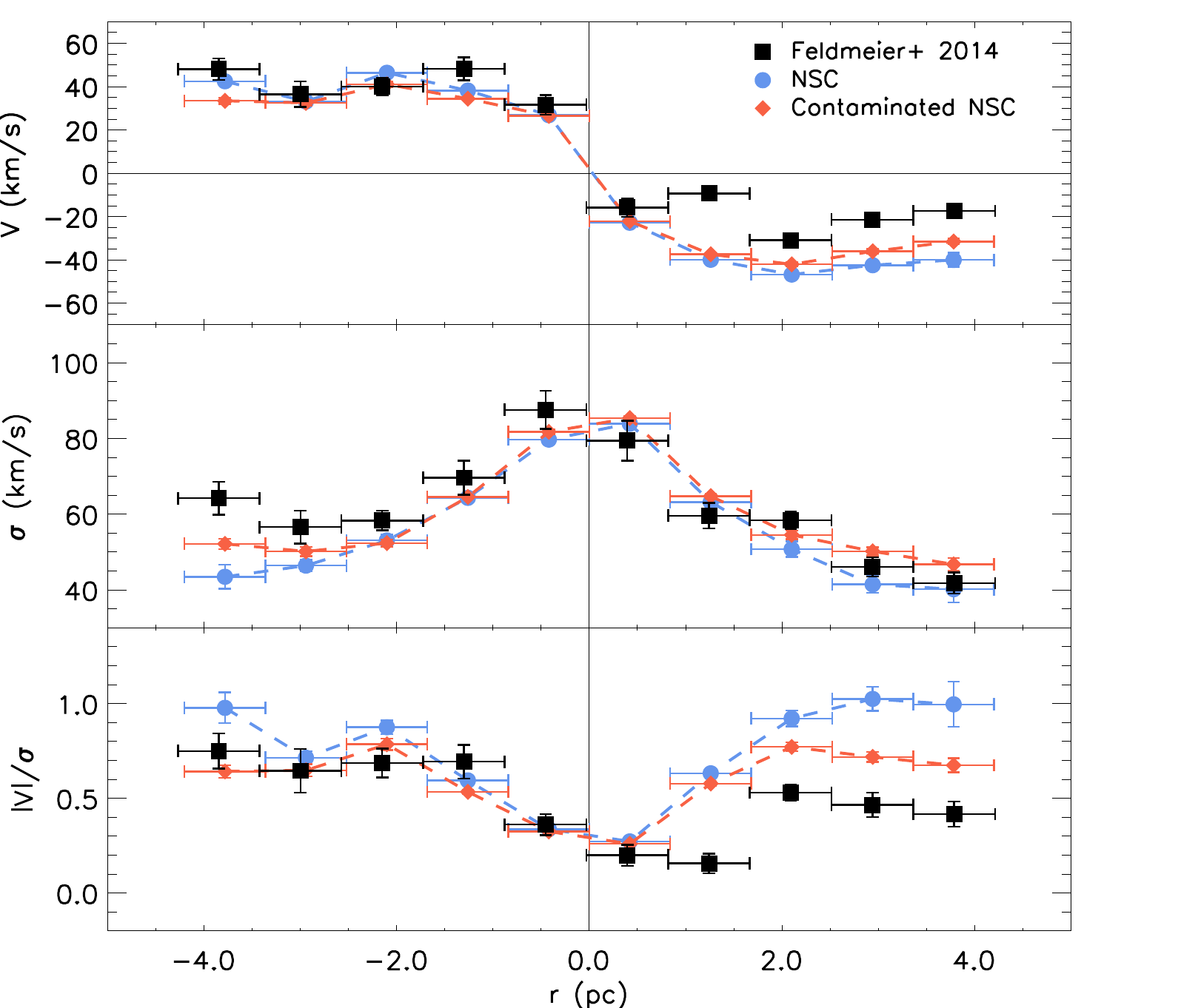}
\center
\caption{Same as Figure~\ref{fig:figure4} for the NSC of Simulation III (blue circles), but accounting for contamination from stars of the surrounding nuclear bulge (red diamonds).}
\label{fig:figure8}
\end{figure}

\subsection{Kinematic substructures}

A kinematic substructure, most likely associated with an old stellar population, has been found at the centre of our MW NSC \citep{Feldmeier_2014}. This substructure is almost aligned with the Galactic minor axis at a distance of 0.8 pc from the Galactic centre. 

To quantify the ability of cluster-inspirals on creating such kinematic substructures, we create density maps of the angular momentum vector distribution for all our simulated NSCs (Figure~\ref{fig:figure5}) within a FoV of 10$\times$10 pc$^2$.
The angular momentum vector of every particle is projected onto the angle space ($\theta$,$\phi$) of a spherical coordinate system. The z-axis of this system is the rotation axis of the main body of the NSC within 5 pc, and $\theta$ and $\phi$ are the polar and the azimuthal angles, respectively, of the angular momentum vector of every particle.

In this space the region $\theta\in[0^{\circ}, 45^{\circ})$ corresponds to prograde (direct) rotating particles (D), $\theta\in[135^{\circ}, 180^{\circ})$ to retrograde (R) and $\theta\in[45^{\circ}, 135^{\circ})$ to particles rotating around the major axis or polar (P; where P+ has $\phi\textless180^{\circ}$ and P- has $\phi\textgreater180^{\circ}$, and correspond to particles with opposite polar spins).

We see that in Simulation III most of the particles end up with prograde rotation ($\theta\textless45^{\circ}$), while in Simulations I and II there is a significant fraction of angular momentum with $\theta\textgreater45^{\circ}$ (polar or retrograde rotation). Both Simulation I and II show strong kinematic substructures up to the 9th infall. However, as the mass ratio between each infalling GC and the NSC keeps decreasing, the GCs have less of an impact on the central kinematics of the NSC. 

In order to translate this to observable properties, we create mock line-of-sight velocity maps of all the simulated NSCs at the last timesteps of their evolution and apply the kinemetry method \citep{Krajnovic_2006}. Figure~\ref{fig:figure6} shows one of these kinematic maps, that corresponds to the NSC of Simulation II, after the 12th infall. The kinemetric model shows a weak polar kinematic twist, resulting from an imbalance between particles with opposite polar spins P+ and P- in Figure~\ref{fig:figure5}. This apparent substructure is not observed, however, in the next (last) timestep, after the NSC evolves for $\sim$2 Gyr in isolation.

\cite{Cole_2016} suggested that it is observationally difficult to kinematically distinguish stars belonging to a specific GC progenitor in the NSC remnant, except in the case where the GC is the most recently (last) accreted. This is also confirmed in our simulations. The stars responsible for the projected polar twist belong to all the four previously accreted GCs. Although in our case the last accreted GCs do not affect significantly the central kinematics of the NSC, we show that the existence of a projected central kinematic twist is a possible outcome of the process of phase space mixing of the stellar populations belonging to different GCs.

We conclude that in our adopted models the NSC cannot exhibit a strong kinematically distinct component in the case where the infalling GCs inspiral with a similar orbital spin (Simulation III). In the case that the GCs infall with random orbital directions, however, the NSC ends up with a significant amount of non-regular rotation, which can translate into weak substructures in the projected kinematics. We note that these substructures could be enhanced with a different choice of initial orbital parameters or structural properties of the infalling GCs. Such a study is, however, beyond the scope of this paper.

\section{The effect of stellar foreground contamination}
\label{sec:Contamination}

Integral Field Unit (IFU) observations of NSCs are naturally contaminated by non-NSC stellar populations of the host galaxy along the line-of-sight. We study the effect of contamination from non-cluster stars to the observed LOSVD of NSCs. We construct mock kinematic maps adopting the same technique as described in Section~\ref{sec:mock_kin}, accounting for the contribution from all the surrounding nuclear bulge stars in our simulations that are present along the line-of-sight and within the adopted field-of-view of 10$\times$10 pc$^2$. The resulting LOSVDs of all the simulation particles (NSC and nuclear bulge) are mass-weighted and fitted using Gauss-Hermite series, as done for the non-contaminated case.

The LOSVD of the ``contaminated" NSCs is shown in Figure~\ref{fig:figure7}. The nuclear bulge stars affect mostly the outer parts of the projected kinematics, where the mass density of the NSC drops and the surrounding nuclear bulge starts to dominate. The nuclear bulge, being non-rotating and dynamically ``hotter'' than the NSC, is causing the LOS velocity to decrease and the velocity dispersion to increase in the outer parts. The higher order moments are also affected-- $h_3$ slightly decreases, while $h_4$ increases in the outer parts. This increase of $\sigma$ and $h_4$ in the outer parts serves as a signature of the existence of contaminating nuclear bulge stars in the LOS kinematics of a NSC.

Accounting for this contamination, the model that reproduces best the observations of the MW NSC is the one that shows the highest intrinsic rotation, resulting from Simulation III (Figure~\ref{fig:figure8}).

We should note, however, that the amount of rotation in our contaminated NSC yields a lower limit to the rotation that would be observed if the NSC was embedded in a more realistic galactic environment. This is due to the fact that our contaminant is a dynamically hot, non-rotating component, while in reality the contamination along the line-of-sight can also contain rotating components (e.g. stars from the Galactic disc and bar) that would affect the observed LOSVD of the NSC.

\section{Conclusions and Discussion}
\label{sec:Summary}

\subsection{Summary}
In this paper we explore whether and how the cluster-inspiral formation scenario can account for the observed rotation and kinematic properties of NSCs, focusing in particular to the MW NSC. We use N-body simulations of the consecutive infall of globular clusters (GCs) in the centre of a MW-like nucleus and construct mock line-of-sight kinematics of the resulting NSCs. Our results can be summarised as follows:

(i) We find that NSCs formed through GC inspirals can show a significant amount of rotation, even if the GCs are initially randomly distributed around the centre. We conclude that no fine tuning of the orientation of the inspiralling GCs is needed to result in a rotating NSC.

(ii) Both the flattening and the kinematic properties of our simulated NSCs match the observed properties of the MW NSC very well.

(iii) In the case where the GCs fall into the centre from random directions, the resulting NSC shows a significant amount of non-regular motion, which can result in projected kinematic misalignments and weak kinematic twists. In the case that the GCs fall in with a similar orbital orientation (e.g. if they originate from the Galactic disc), the resulting NSC shows more rapid and regular rotation.

(iv) Given that IFU observations are naturally contaminated with stars from the nuclear bulge surrounding the NSC, we find that such a contamination lowers the observed rotation and imprints its dynamical signature in the outer parts (r$\textgreater$2pc), accounting for an increase in the LOS velocity dispersion $\sigma$ and the kurtosis-like higher-order moment $h_4$ of the NSC.

\subsection{Discussion}

We have studied the formation of NSCs solely through cluster-inspirals and do not exclude the possibility that gas accretion and in situ star formation play a role in their formation. The prevalence, if any, of each formation mechanism should be connected to the galactic environment of the NSC. 

In that sense, the cluster-inspiral scenario is expected to play a dominant role in the formation of rotating NSCs in early-type galaxies that are too gas-poor to support the formation of a NSC solely through gas accretion. However, the NSC formation could have happened at a time when gas was still present. Observational evidence seem to support the cluster-inspiral scenario especially for low-mass early-type galaxies \citep[e.g.][]{Turner_2012, denBrok_2014}. High-mass early-type galaxies seem to show a more complex nature: the existence of counter-rotating populations in NSCs points towards the cluster-inspiral scenario, while the complex stellar populations that they host point towards episodic gas accretion and in situ star formation \citep{Lyubenova_2013}.

In the case of late-type host galaxies such as our Milky Way, both mechanisms are expected to work in parallel, as supported by their observed metallicity spreads and complex star formation histories \citep[e.g.][]{Rossa_2006, Do_2015}. However, the contribution of each mechanism to the main build-up process of the MW NSC is not clear. Observational evidence shows that there has been an increase in star formation in the last few hundred Myr of evolution of our MW NSC \citep{Blum_2003, Pfuhl_2011}. In the very centre ($\sim$0.5 pc) the light is dominated by 6 Myr old stars \citep[e.g.][]{Paumard_2006, Feldmeier_2015} and in the central 1pc only a small fraction of low-metallicity stars are consistent with the typical metallicities of MW globular clusters \citep{Do_2015}, favoring in situ star formation from gas accretion.

On the other hand, approximately $\sim80\%$ of the MW's NSC stars in the central $\sim$2.5 pc were formed more than 5 Gyr ago \citep{Blum_2003, Pfuhl_2011}. Kinematic evidence (e.g. the centrally peaked $V_{RMS}$, the kinematic offset from the Galactic disc, the evidence for a polar kinematic substructure) support the cluster-inspiral formation scenario. Here we have shown that cluster-inspirals can also account for the observed rotation of NSCs, an evidence that has been so far attributed to gas infall.

The search for the dominant formation mechanism of NSCs is still ongoing and its connection to their galactic environment seems far from a foregone conclusion. Clarifying the nature of NSC formation would now require more detailed studies of their dynamics, their stellar populations and star formation history, combined with more realistic simulations of their formation.


\section*{Acknowledgements}

We are grateful to Anja Feldmeier-Krause and Paola Di Matteo for their useful comments and contribution to this work. AT and GvdV acknowledge financial support to the DAGAL network from the People Programme (Marie Curie Actions) of the European Union's Seventh Framework Programme FP7/2007- 2013/ under REA grant agreement number PITN-GA-2011-289313. AT and PB acknowledge support from the International Max-Planck Research School for Astronomy and Cosmic Physics at the University of Heidelberg (IMPRS-HD). HBP acknowledges support from the Israeli I-CORE center for astrophysics under ISF grant 1829/12.




\bibliographystyle{mnras}








\bsp	
\label{lastpage}
\end{document}